\documentstyle[12pt]{article}

\begin{document}
\begin{titlepage}
\vspace{2.0cm}
\begin{centering}
{\LARGE{\bf Closed Form Results for Shape Transitions in Lipid Monolayer Domains}}\\
\bigskip\bigskip
Jos\'e A. Miranda\footnote{e-mail:jme@lftc.ufpe.br}\\
{\em Laborat\'orio de F\'{\i}sica Te\'orica e Computacional, Departamento de F\'{\i}sica\\
Universidade Federal de Pernambuco\\ 
Recife, Pernambuco  CEP. 50670-901 Brazil}\\
\vspace{0.5cm}

\end{centering}

\begin{abstract}

Isolated domains in lipid monolayers exhibit shape 
transitions as a consequence of the competing effects of line tension 
and long-range dipolar interactions. Nontrivial integrals ultimately 
arise in stability analysis calculations of these domains. In this work 
we present a closed form evaluation of such integrals in terms of 
associated Legendre functions. Our approach is motivated by and extends 
the work of J. M. Deutch and F. E. Low (J. Phys. Chem. {\bf 1992}, 96, 
7097-7101). Our closed form solutions lead to simpler analytic expressions 
for stability thresholds, which are easier to evaluate explicitly. 
Relevant asymptotic behavior is calculated in various limits of interest. 
For the case of Coulomb interactions, a general closed form expression 
is obtained for the critical domain radii.

\end{abstract}

\hspace{0.8 cm}
\end{titlepage}
\def\carre{\vbox{\hrule\hbox{\vrule\kern 3pt
\vbox{\kern 3pt\kern 3pt}\kern 3pt\vrule}\hrule}}

\baselineskip = 30pt

\section{Introduction}
In the past few years the study of shape transitions in lipid 
monolayer domains at the air-water interface has received considerable 
attention from the physical-chemical community~\cite{Rev}. 
The great interest towards this subject is in 
part due to the diversity of domain shapes, which range from simple 
circular structures to complex labyrinthine patterns. Theoretical 
investigations study this variety of shapes by a model in 
which an array of molecular dipoles is arranged normal to the 
two-dimensional domain. The basic mechanism behind the formation 
of such structures arises from the 
competition between interfacial tension and long-range electrostatic 
interactions. Line tension acts to minimize the domain perimeter and 
favors compact domain shapes. In contrast, repulsive electrostatic 
dipole-dipole interaction tends to maximize the distance between dipoles 
and favors extented domain shapes. In this context, the total area of 
the domain is not changed.

In previous work McConnell~\cite{McC}, Deutch and Low~\cite{Deu}, 
and Goldstein and Jackson~\cite{Gol} have studied the 
shape transitions of initially circular (radius $R$) 
lipid monolayer domains in a lowest order (quadratic) 
stability analysis. In all these theoretical studies many 
relevant quantities are calculated exactly, but due to their complexity, 
some of the computed expressions are not evaluated in closed form. 
For example, the formulas for stability thresholds found 
in~\cite{McC,Deu,Gol} are usually expressed in terms of nontrivial 
integrals. Related integrals arise in stability investigations 
of other systems, also presenting polarized domains, such as 
planar-confined ferrofluids in magnetic fields~\cite{Lan} and 
thin ferromagnetic films forming magnetic bubble patterns~\cite{Thi}.

A considerable difficulty to perform stability analysis of lipid monolayer 
domains is the evaluation of the electrostatic dipole-dipole energies 
for various domain shapes. These calculations involve the manipulation 
of double line integrals which are obtained by an 
ingenious application of Green's theorem~\cite{McC4}. The subtleties 
concerning the validity of Green's theorem under such circumstances are 
connected to the necessity of introducing a cutoff distance required to 
prevent divergence at zero separation between dipoles~\cite{Comment3}. 
As shown by Deutch and Low~\cite{Deu}, the subtle 
application of Green's theorem 
can be avoided by calculating the electrostatic energy directly through 
area integrals.

The purpose of the present work is to evaluate in closed form the integrals 
and pertinent stability conditions obtained in reference~\cite{Deu}. Even 
though the nontrivial integrals which appear in references~\cite{McC,Deu,Gol} 
basically belong to the same general family, we focus on Deutch and Low's paper~\cite{Deu} since their work precisely illustrates the utility of 
our closed form expressions. Our calculations are performed in the 
same spirit of those by Miranda and Widom~\cite{Mir}, 
who recently carried out a related closed form study focused on ferrofluids. 
We evaluate the integrals in terms of associated Legendre 
functions. Asymptotic behavior in interesting limits such as small 
cutoff, or high spatial frequency 
are rigorously calculated. Finally, we address the case which considers 
repulsive Coulombic interactions within a charged domain. 
Comparison between our results and those derived in~\cite{Deu} is 
consistently made throughout the text. We obtain simpler 
and compact expressions which facilitate the explicit evaluation 
of many important quantities derived in previous studies of 
shape transitions in lipid monolayers~\cite{McC,Deu,Gol}.

\section{Stability analysis of lipid monolayer domains - Closed form results}

\subsection{The integrals $B_{n}(\hat{\alpha})$}

Following Deutch and Low~\cite{Deu}, the perturbed domain shape is described 
by the Fourier expansion
\begin{equation}
\label{interface} 
r(\theta)= R + \sum_{n \ne 0} \zeta_{n} \exp{(i n \theta)}, 
\end{equation}
where $\theta$ is the polar angle, $n$ (an integer) denotes the discrete azimuthal wave number and $\zeta_{n}$ represents the perturbation 
amplitudes (figure 1). In equation~(\ref{interface}) 
$\zeta_{0}$ is expressed in terms of the amplitudes $\zeta_{n}$ 
$(n \ne 0)$ in order to satisfy the condition that the total 
area ${\cal A}=(1/2) \int_{0}^{2 \pi} r(\theta)^{2} d\theta$ 
of the domain remains constant and independent of the 
perturbation, i.e., ${\cal A}=\pi R^{2}$. The 
perturbation mode with $|n|=1$ corresponds to a global off-center 
shift (rigid translation) of the unperturbed initial shape. It is a 
spurious, shape-preserving mode that does not contribute to domain 
energy changes.

The total energy of the domain described by $r(\theta)$ can be 
calculated by considering electrostatic and line-tension contributions. 
The bulk (dipole-dipole) electrostatic 
energy of a distribution of molecular dipoles oriented in a direction 
normal to the two-dimensional domain is given by the double surface integral
\begin{equation}
\label{bulk}
W[r(\theta)]=\frac{\mu^2}{2} \int d^{2}r \int d^{2}r' \frac{1}{[|\vec r - \vec {r}'|^2 + \alpha^2]^{3/2}},
\end{equation}
where $\mu$ is the dipole density of the surface and $\alpha$ is a cutoff 
distance of closest approach of neighboring dipoles. On the other hand, the line-tension energy, given by the product of line tension $\lambda$ 
by the domain perimeter, is
\begin{equation}
\label{line}
F_{\lambda}[r(\theta)]= \lambda \int_{0}^{2 \pi} d\theta 
\left [r(\theta)^2 + \left ({dr(\theta) \over d\theta} \right)^2 \right ]^{1/2}.
\end{equation}

By using equations~(\ref{interface})-(\ref{line}), Deutch and Low~\cite{Deu} 
showed that the total energy $F[r(\theta)]=W[r(\theta)] + 
F_{\lambda}[r(\theta)]$, accurate to order ${\cal O}(\zeta_{n}^2)$, can be 
expressed in the form
\begin{equation}
\label{total}
F=F_{0} + \frac{2 \pi}{R} \sum_{n > 0} \Omega_{n} |\zeta_{n}|^2,
\end{equation}
where $F_{0}$ is the energy of the unperturbed circle and 
\begin{equation}
\label{omega}
\Omega_{n}=\lambda(n^2 - 1) - 2 \mu^2 [B_{1}(\hat{\alpha}) - B_{n}(\hat{\alpha})],
\end{equation}
where
\begin{equation}
\label{Bn}
B_{n}(\hat{\alpha})=\frac{1}{2^{3/2}}\int_{0}^{\pi} \frac{\cos{n\theta}}{\left [ \left ( 1 + \frac{\hat{\alpha}^{2}}{2} \right) - \cos{\theta}\right]^{3/2}} d\theta
\end{equation}
and $\hat{\alpha}=\alpha/R$.
We can examine the stability of the circle simply by 
looking at the sign of $\Omega_{n}$: instability occurs when $\Omega_{n} < 0$. 
Similar results have been obtained in references~\cite{McC} 
and~\cite{Gol}, despite subtleties regarding their implementation 
of the cutoff. Even though expression~(\ref{omega}) is exact 
(to second order), it is written in terms of the integrals 
$B_{n}(\hat{\alpha})$, and have not been explicitly evaluated. 
Consequently, relevant physical quantities like 
the critical radii $R_{n}$ (radii beyond which a circular domain becomes 
unstable with respect to $n$-fold harmonic perturbations) are often written 
in an indirect fashion, which involves complicated sums and integrals, 
or requires the use of recursion relations~\cite{comment}. The lack of 
closed form results makes difficult the precise and explicit evaluation 
of $R_{n}$ for arbitrary $n$. Therefore, it is of practical 
interest to find closed form solutions for the integrals $B_{n}(\hat{\alpha})$.

The closed form expression for the integrals $B_{n}(\hat{\alpha})$ can be 
found with the help of the integral representation for the 
associated Legendre function of the second kind~\cite{Mag} 
\begin{eqnarray}
\label{associated}
Q_{\nu}^{m}(z) & = &\frac{\exp(m \pi i)}{(2 \pi)^{1/2}} \Gamma{ \left (m + \frac{1}{2} \right )} (z^2 - 1)^{m/2} \Bigg \{ \int_{0}^{\pi} \frac{\cos \left (\nu + \frac{1}{2} \right ) t}{[z - \cos t]^{m + 1/2}} dt \nonumber \\
               & - &\cos(\nu \pi) \int_{0}^{\infty} \frac{\exp \left [- \left ( \nu + \frac{1}{2} \right ) t \right ]}{[z + \cosh t]^{m + 1/2}} dt \Bigg \}, 
\\ \nonumber
\end{eqnarray}
with ${\rm Re}(m) > -1/2$, ${\rm Re}(\nu + m)>-1$ and $|{\rm arg}(z \pm 1)| < \pi$. 
$\Gamma$ denotes the 
gamma function. 
Set $\nu=n - 1/2$ and $m=1$ in equation~(\ref{associated}) and 
compare the resulting expression with equation~(\ref{Bn}) to get
\begin{equation}
\label{closedBn}
B_{n}(\hat{\alpha})= - \frac{ Q_{n - 1/2}^{1} \left( 1 + \frac{\hat{\alpha}^{2}}{2} \right)}{\hat{\alpha} \left (1 + \frac{\hat{\alpha}^{2}}{4} \right )^{1/2} }.
\end{equation}
This way, equation~(\ref{omega}) can be 
rewritten as
\begin{equation}
\label{omega2}
\Omega_{n}=\lambda(n^2 - 1) - 2 \mu^2 \left [ \frac{Q_{n - 1/2}^{1} \left( 1 + \frac{\hat{\alpha}^{2}}{2} \right) - Q_{1/2}^{1} \left( 1 + \frac{\hat{\alpha}^{2}}{2} \right)}{\hat{\alpha} \left (1 + \frac{\hat{\alpha}^{2}}{4} \right )^{1/2}} \right ].
\end{equation}
We stress that $\Omega_{n}$ is now written in a closed form fashion, which 
explicitly expresses its precise functional dependence on $n$, $\hat{\alpha}$, 
$\lambda$ and $\mu$. The fact that $\Omega_{n}$ is written in terms of 
the associated Legendre functions is very convenient, since their 
fundamental properties and basic functional behavior are 
quite well known~\cite{Mag}. The usefulness of equation~(\ref{omega2}) 
becomes specially evident in the calculation of asymptotic 
expressions in important limiting situations 
such as small cutoff ($\alpha$ $\rightarrow 0$) and long wavelengths 
($n$ $\rightarrow \infty$, $\alpha$ $\rightarrow 0$, with $\hat{\alpha}$ kept 
finite and small). In such relevant limits our closed solution 
leads to simple and elegant expressions, which are easier to 
evaluate than those derived in~\cite{Deu}.

In order to gain insight about the role of $\Omega_{n}$ in 
the domain instability, we plot in figure 2 
the ratio $\Omega_{n}/2\lambda$ as a function of mode number $n$. 
We assume that $\hat{\alpha}=2 \times 10^{-5}$ and take three different 
values of the dimensionless parameter $\mu^{2}/\lambda$, a quantity that 
characterizes the relative importance of dipolar and surface energies. 
All points are obtained exactly, using our equation~(\ref{omega2}). 
The domain boundary remains stable to $n$-th mode distortions up to a 
critical value $D_{Cr}(n)$ of $\mu^{2}/\lambda$, defined by setting 
$\Omega_{n}=0$ in equation~(\ref{omega2}). In (a) $\mu^{2}/\lambda < 
D_{Cr}(2)$, so $\Omega_{n} \ge 0$ and the domain is stable. For larger 
values of $\mu^{2}/\lambda$ (see (b) and (c)) $\Omega_{n}$ can be negative 
and the domain deforms.

\subsection{Stability condition and critical radii $R_{n}$}

For amphiphilic systems the ratio between the microscopic cutoff $\alpha$ and 
the radius $R$ of the circular domain is a small quantity. In this section, 
we take advantage of our closed form expression~(\ref{omega2}) to evaluate 
the stability condition and critical radii $R_{n}$ in the limit of 
small $\hat{\alpha}$.

From our discussion in the previous section the stability 
condition $\Omega_{n} \ge 0$ for 
circular lipid monolayer domains can be written as
\begin{equation}
\label{stability1}
\frac{\lambda}{\mu^{2}} \ge \frac{2}{(n^{2} - 1)} \left [ \frac{Q_{n - 1/2}^{1} \left( 1 + \frac{\hat{\alpha}^{2}}{2} \right) - Q_{1/2}^{1} \left( 1 + \frac{\hat{\alpha}^{2}}{2} \right)}{\hat{\alpha} \left (1 + \frac{\hat{\alpha}^{2}}{4} \right )^{1/2}} \right ].
\end{equation}
In order to evaluate~(\ref{stability1}) in the limit of small $\hat{\alpha}$, 
we use the following property of the associated Legendre functions~\cite{Mag}
\begin{equation}
\label{property}
Q_{\nu}^{m}(z)= (z^{2} - 1)^{m/2} {d^{m}Q_{\nu}(z) \over dz^{m}},
\end{equation}
to rewrite~(\ref{stability1}) as 
\begin{equation}
\label{stability2}
\frac{\lambda}{\mu^{2}} \ge \frac{{d \over dx} \left [Q_{n - 1/2}(1 + 2x) - Q_{1/2}(1 + 2x)\right ]}{(n^{2} - 1)},
\end{equation}
where $x=\hat{\alpha}^2/4$. 

To get a closed form expression for the stability 
condition~(\ref{stability2}) we need to evaluate 
$Q_{n - 1/2}(1 + 2x)$ for small values of $x$. 
To succeed in doing this, we use reference~\cite{Rob} to 
expand $Q_{n - 1/2}(1 + 2x)$ near $x=0$, to first order in $x$
\begin{eqnarray}
\label{smallx}
Q_{n - 1/2}(1 + 2x) & \approx & \left [ -C - \frac{1}{2} 
\ln\left (\frac{x}{x + 1} \right ) 
- \psi\left ( n + \frac{1}{2} \right ) \right ] + \nonumber\\
& & \left [ (1 - C) - \frac{1}{2} \ln\left (\frac{x}{x + 1} \right ) 
- \psi\left ( n + \frac{1}{2} \right ) \right ] 
\left (\frac{4n^{2} - 1}{4} \right )x \nonumber \\ 
& & + \, {\cal O}(x^{2} \ln x, x^{2}),
\end{eqnarray}    
where Euler's psi function $\psi$ is the logarithmic derivative of the gamma function~\cite{Mag} and $C$ is Euler's constant. 

Using equations~(\ref{stability2}) and~(\ref{smallx}), we obtain the 
closed form stability condition, accurate to order $\alpha$,
\begin{equation}
\label{small-alpha}
\frac{\lambda}{\mu^{2}} \ge \ln\left (\frac{8R}{\alpha e} \right ) - \left \{ \frac{1}{4} \left ( \frac{4n^{2} - 1}{n^{2} - 1} \right ) \left [ \psi\left (n + \frac{1}{2} \right ) - \psi\left (\frac{3}{2} \right ) \right ] + \frac{1}{2} \right \}.
\end{equation}
At this point, we compare our expression~(\ref{small-alpha}) with an equivalent 
result obtained by Deutch and Low~\cite{Deu} (equation (36) in their work). 
In reference~\cite{Deu} the part corresponding to the term 
between curly brackets on the right hand 
side of~(\ref{small-alpha}) is written as complicated 
definite integrals involving Chebyshev polynomials. The authors in~\cite{Deu} 
do not solve such integrals explicitly. In contrast, our 
derivation, based on the well known properties of the associated 
Legendre functions, naturally provides a closed form stability condition. 
Obviously, our solutions are useful only if they lead to easier 
manipulation of the obtained results. This is exactly the case: the 
right hand side of our stability condition~(\ref{small-alpha}) 
is much easier to be evaluated explicitly than the equivalent piece 
which appears in~\cite{Deu}. Notice that analytic 
evaluation of equation~(\ref{small-alpha}) can be promptly performed, 
for specific $n$, by means of a simple 
property relating $\psi$ functions~\cite{Mag}
\begin{equation}
\label{property-psi}
\psi\left( n + \frac{1}{2} \right) - \psi\left( n - \frac{1}{2} \right)= \frac{2}{2n - 1}.
\end{equation}

Another important quantity that can be immediatly 
obtained from~(\ref{small-alpha}) refers to the critical radii 
at which the circular domain becomes unstable with respect to a given 
Fourier mode $n$
\begin{equation}
\label{Rn}
R_{n}= \frac{\alpha}{8}\exp\left (\frac{\lambda}{\mu^{2}} \right ) \exp{(Z_{n})},
\end{equation}
where the critical paramenter for transition is given by
\begin{equation}
\label{Zn}
Z_{n}=\frac{3}{2} + \frac{1}{4} \left ( \frac{4n^{2} - 1}{n^{2} - 1} \right ) \left [ \psi\left (n + \frac{1}{2} \right ) - \psi\left (\frac{3}{2}\right) \right ].
\end{equation}
The values calculated for $Z_{n}$ directly using our equation~(\ref{Zn}) 
agree precisely with those computed in equation (39) of 
reference~\cite{Deu}. Again, the major difference between our 
results resides on the fact that in~\cite{Deu}, 
to get a particular value of $Z_{n}$ it is necessary to appeal to 
an indirect approach which requires solving several different integrals 
involving Chebyshev polynomials, for each $n$ to be considered. 
Our equation~(\ref{Zn}) offers a simpler, direct and more convenient way 
to evaluate the various $Z_{n}$. 

Figure 3 depicts the dimensionless critical radius 
$R_{n}/\alpha$ plotted against the ratio $\lambda/\mu^{2}$. 
All curves are computed analytically from our equations~(\ref{Rn}) 
and~(\ref{Zn}). This figure illustrates important physical behavior 
captured by lowest order stability analysis. 
Mode $n=2$ is always the first to become 
unstable (elliptical instability), establishing a stability boundary. 
At fixed value of $\lambda/\mu^{2}$, higher modes $(n \ge 3)$ become 
unstable at larger $R_{n}/\alpha$. As $R_{n}/\alpha$ increases, each 
mode becomes dominant for a range of radii, and hence there is a cascade 
to higher modes. For a given critical radius, 
smaller values of $\lambda/\mu^{2}$ lead to stronger mode competition, 
meaning more intrincated domain boundaries. Related stability 
diagrams have been numerically obtained in~\cite{Lan} and~\cite{Lee}. 
The critical parameter $Z_{n}$ is plotted as a function of mode number 
$n$ in the inset of figure 3.

\subsection{Relevant asymptotic behavior and the Coulomb interaction case}
  
We begin this section calculating further asymptotic expansions
by taking advantage of our closed form expression~(\ref{stability1}). 
First, consider the large $n$ and small $\alpha$ limit, assuming that 
the product $n\hat{\alpha}$ is kept fixed. This is the limit of interest 
in many practical situations~\cite{McC,Deu,Gol}. 
Considering the large $n$ limit, 
use the asymptotic expression~\cite{Mag}
\begin{equation}
\lim_{n \rightarrow \infty} \frac {Q_{n - 1/2}^{m}\left [\cosh \left( \frac{\tau}{n} \right) \right ]}{n^{m} \exp(i \pi m)}=K_{m}(\tau), 
\end{equation}
where $K_{m}(\tau)$ is the modified Bessel function of order $m$, and set 
\begin{equation}
\tau=n \cosh^{-1} \left( 1 + \frac{\hat{\alpha}^{2}}{2} \right ) \approx n\hat{\alpha} + {\cal O} (n\hat{\alpha}^{3}), 
\end{equation} 
to rewrite stability condition~(\ref{stability1}) as
\begin{equation}
\label{largen}
\frac{\lambda}{\mu^{2}} \ge \frac{2}{n^{2}} \left [ \frac{ K_{1}(\hat{\alpha}) - n K_{1}(n\hat{\alpha})}{\hat{\alpha} \left (1 + \frac{\hat{\alpha}^{2}}{4} \right )^{1/2}} \right ].
\end{equation}
In addition, take into account the fact that $\hat{\alpha}$ is small, 
keeping $n\hat{\alpha}$ fixed, to obtain
\begin{equation}
\label{largen-smallalpha}
\frac{\lambda}{\mu^{2}} \ge 2 \left [ \frac{1}{(n\hat{\alpha})^{2}} - \frac{1}{n\hat{\alpha}}K_{1}(n\hat{\alpha}) \right ],
\end{equation}
where we have used the small argument expansion
\begin{equation}
\label{Besselexpansion}
K_{1}(y) \approx \frac{1}{y} + \frac{1}{2} \left [ \left (C - \frac{1}{2} \right ) + \ln \left ( \frac{y}{2} \right ) \right ]y + {\cal O}(y^{3} \ln y, y^{3})
\end{equation}
and a power series expansion of the denominator in~(\ref{largen}).
Equation~(\ref{largen-smallalpha}) is quite general, being valid 
for large values of $n$ and small values of $\alpha$, provided 
$n\hat{\alpha}$ is kept finite. We would like to 
point out that our result~(\ref{largen-smallalpha}) differs from 
the equivalent one obtained in~\cite{Deu} (see equation (46) in 
their work). Because of a mistake in their calculation, Deutch and Low 
have written their equation (46) in terms of the modified Bessel 
function of order zero $K_{0}(n\hat{\alpha})$ 
(there is also a sign error), which is 
not the correct result. This minor mistake would be of no consequence if 
the authors in~\cite{Deu} would have not derived an important subsequent 
result, using their incorrect expression (46) as a starting point. 
This fact generates some confusion. Further 
comments on this issue are given in the next paragraph.

We take advantage of our equation~(\ref{largen-smallalpha}) to discuss 
the so-called straight edge limit\cite{Deu,McC2}. This is the situation 
in which the domain interface is flat with a small perturbation 
of wave number $k$. To get this limit from our circular geometry, 
we take radius $R \rightarrow \infty$ at fixed $\alpha$. 
Fixed wave number $k \equiv n/R$ implies $n \rightarrow 
\infty$ and $\hat{\alpha} \rightarrow 0$, with fixed $n\hat{\alpha}$ 
as in our asymptotic expansion equation~(\ref{largen-smallalpha}). 
Taking the limit of~(\ref{largen-smallalpha}) for large wavelengths 
compared to the cutoff ($k\alpha \ll 1$), we use the 
small argument expansion~(\ref{Besselexpansion}) 
to get the critical spatial frequency
\begin{equation}
\label{criticalk}
k_{c}=\left [ \frac{\alpha}{2} \exp \left (\frac{\lambda}{\mu^{2}} + C - \frac{1}{2} \right ) \right ]^{-1}.
\end{equation}
The straight edge of the domain is unstable for all spatial 
frequencies $k < k_{c}$. This result agrees with 
equation (13) in McConnell's work~\cite{McC2} 
since the line tension considered in his derivation 
contains an additional electrostatic contribution 
given by $-\mu^2$~\cite{Comment2}. 
Our result~(\ref{criticalk}) also agrees with the 
quantity $R_{M}(n) \equiv n/k_c$ 
derived by Deutch and Low (equation (48)). 
In principle, we should not expect agreement since $R_{M}(n)$ 
in~\cite{Deu} is obtained using their incorrect intermediate 
expressions (46) and (47), as pointed out in the previous paragraph. 
It is worth mentioning that the critical spatial 
frequency~(\ref{criticalk}) could have been calculated by taking 
the large $n$ limit of our earlier results~(\ref{Rn}) and~(\ref{Zn}).

We conclude by extending our analysis to the case of 
Coulomb interactions in charged domains~\cite{McC3}. 
As indicated in~\cite{Deu}, for a Coulomb interaction between 
two particles (charge $q$) of the form $u(\rho)=(q^{2}/\rho)$, 
with $\rho=[r^{2} + r'^{2} - 2rr'\cos\theta]^{1/2}$, 
the domain stability condition can be written as
\begin{equation}
\label{Coulomb}
\lambda(n^{2} - 1) \ge 2^{1/2} q^{2} R^{2}[{\cal B}_{1}(\theta) - {\cal B}_{n}(\theta)],
\end{equation}
where
\begin{equation}
\label{integral2}
{\cal B}_{n}(\theta)=\int_{0}^{\pi} \frac{\cos{n\theta}}{\left [ 1 - \cos{\theta}\right ]^{1/2}} d\theta.
\end{equation}
As in the case of the dipolar interaction potential, the integrals 
${\cal B}_{n}(\theta)$ are not evaluated in~\cite{Deu}. 
Using equation~(\ref{associated}) with $\nu=n - 1/2$, $m=0$ and 
$z \rightarrow 1$, plus expansion~(\ref{smallx}), we find closed 
form expressions for the integrals ${\cal B}_{n}(\theta)$, and 
consequently for the stability condition~(\ref{Coulomb}). 
With the help of these results, 
we calculate the critical radii for a given $n$ mode distortion
\begin{equation}
\label{criticalradius-C}
R_{c}(n)=\left \{ \frac{\lambda (n^{2} - 1)}{2 q^{2} \left [ \psi \left (n + \frac{1}{2} \right ) - \psi \left (\frac{3}{2} \right ) \right]} \right \}^{1/2}.
\end{equation}
Keller et al.~\cite{McC3} studied the specific case of elliptical 
deformations in charged domains. Our result is identical 
to theirs if we set $n=2$ in equation~(\ref{criticalradius-C}). 
Equation~(\ref{criticalradius-C}) generalizes previous results found 
for $R_{c}(n)$~\cite{Deu,McC3}, providing a closed form expression, 
valid for arbitrary $n$.

\section{Concluding Remarks}

Lipid monolayers form two-dimensional domains as a result 
of the competition between line-tension and long-range repulsive 
interactions. Stability analysis of these domains depends on a 
family of nontrivial integrals. For repulsive electrostatic 
dipole-dipole interaction, we present a closed form evaluation 
of these integrals as a combination of associated Legendre functions. 
Asymptotic behavior in several interesting limits are evaluated. 
A general closed form result is also obtained for the case in 
which the repulsive interaction is of Coulombic nature. 
Our theoretical results should be viewed as complementary 
to the work of Deutch and Low~\cite{Deu}, leading to exact, 
simpler and explicit formulae which commonly appear 
in the study of shape transitions in lipid monolayer domains. 

We hope our results will be helpful for future work on the 
subject of shape instabilities in amphiphilic monolayers.
While the present work has addressed transitions 
associated to a slightly deformed circle, we anticipate that 
the analytical treatment of more complicated domain shapes 
may benefit from our closed form results. In higher 
order perturbative calculations, the various harmonic 
modes couple and cannot be studied separately. In order to 
cope with larger distortions a mode coupling theory is required. 
We point out that a weakly nonlinear analysis 
of such situation can be carried out analytically, following 
a related recent work on viscous fingering instability~\cite{Mir2}. 
Once applied to monolayers, this approach 
would also lead to nontrivial integrals which could be explicitly 
evaluated by employing the methodology and results we presented here.

\pagebreak

\noindent
{\bf Acknowledgments}\\
\noindent
I am grateful to M. Widom for introducing me to the subject 
of pattern formation in polarized domains. I would like to thank 
F. Moraes for his critical reading of the manuscript. It is a pleasure 
to thank A. de P\'adua and M. Oliveira for kindly helping me to 
prepare figure 2. This work was supported by CNPq (Brazilian Agency).

\pagebreak

\noindent
{\Large {\bf Figure Captions}}
\vskip 0.5 in
\noindent
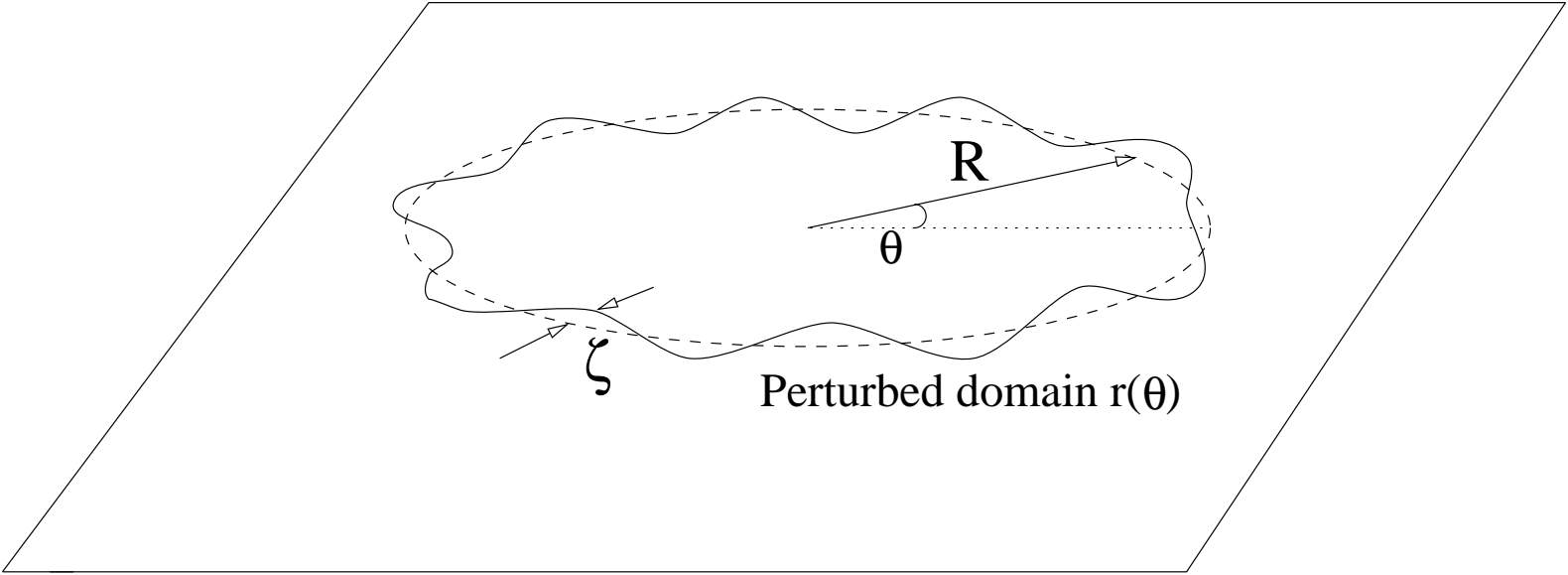
{\bf Figure 1:} Schematic configuration of a two-dimensional 
lipid domain. The dashed line represents an initially circular 
domain of radius $R$ and the solid undulated curve depicts 
its perturbed shape $r(\theta)$= $R + \zeta$. 
The perturbation 
$\zeta$=$\sum\limits_{n \ne 0} \zeta_{n} \exp{(i n \theta)}$, where 
$n$ denotes the azimuthal wave number 
and $\theta$ gives the angular location of the points on the boundary. 
The domain area enclosed by the curve $r(\theta)$ remains constant 
and equals to $\pi R^{2}$.
\vskip 0.25 in
\noindent
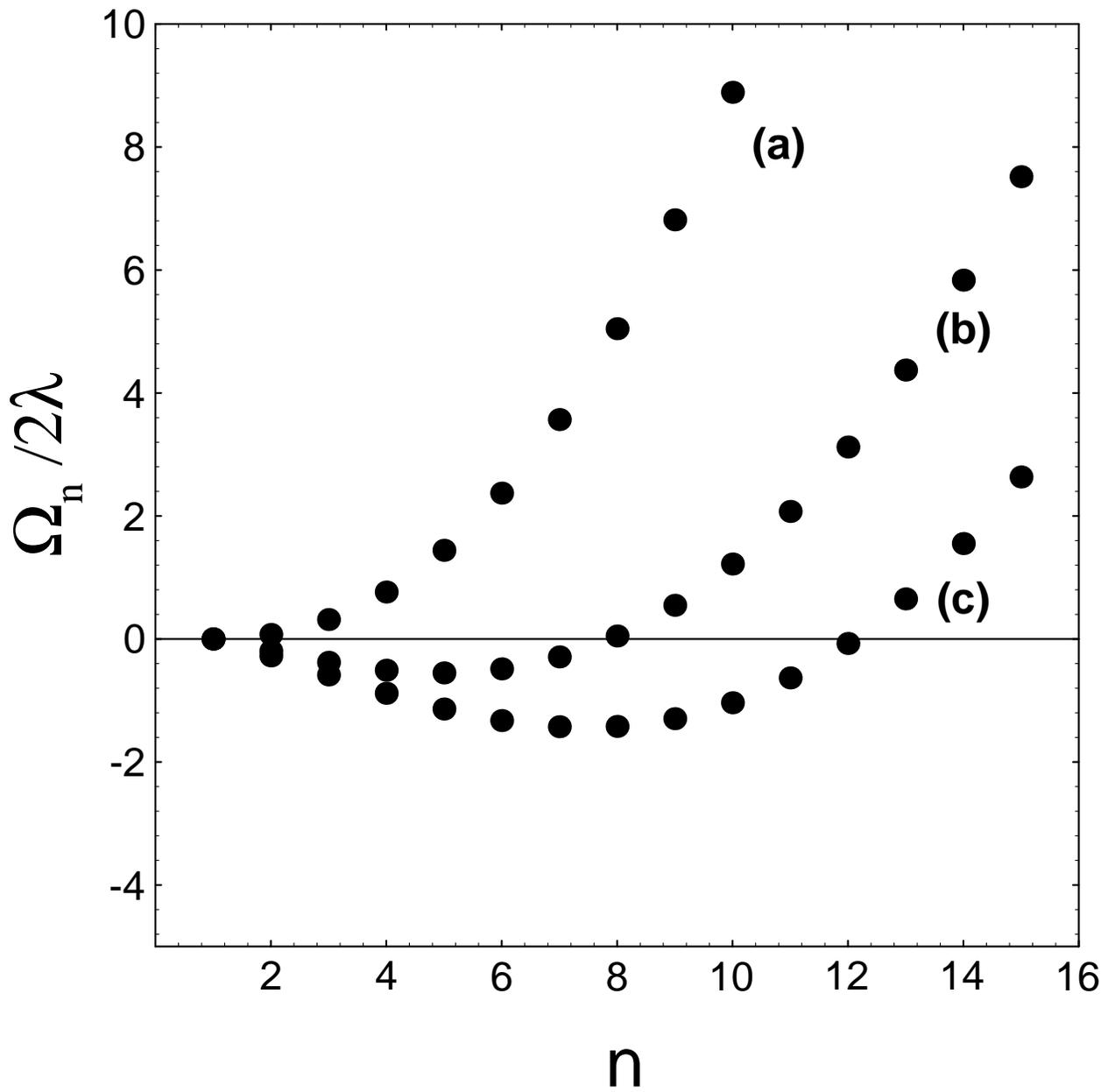
{\bf Figure 2:} Ratio $\Omega_{n}/2\lambda$ as a function of mode number 
$n$, for three distinct values of $\mu^{2}/\lambda$ (a) $0.090$; (b) 
$0.107$; (c) $0.112$. The critical dimensionless parameter $D_{Cr}(2) 
\approx 0.095$ and $\hat{\alpha}=2 \times 10^{-5}$. All points were 
found exactly using our closed form equation~(\ref{omega2}). 
\vskip 0.25 in
\noindent
\vskip 0.25 in
\noindent
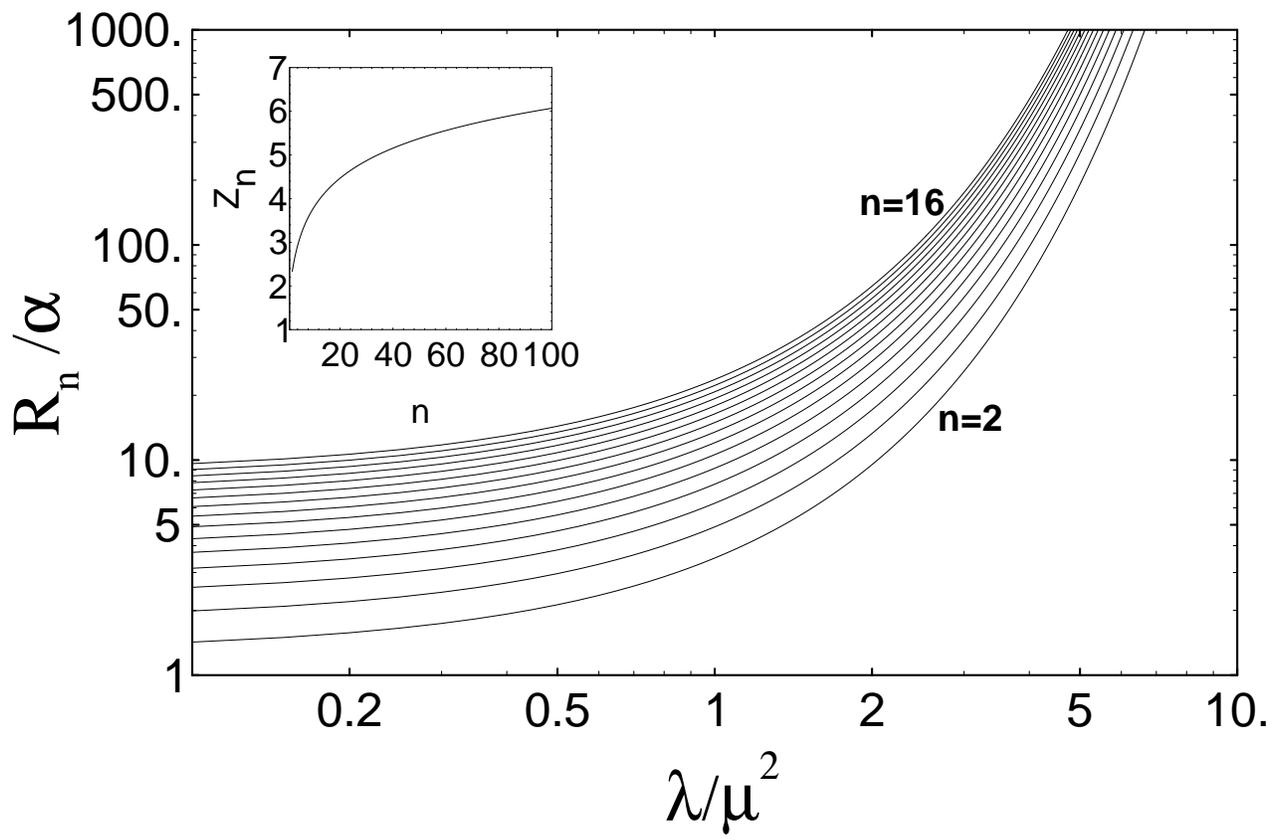
{\bf Figure 3:} Dimensionless critical radius $R_{n}/\alpha$ as a 
function of $\lambda/\mu^{2}$. Closed form equations~(\ref{Rn}) 
and~(\ref{Zn}) were used to plot the various 
curves $2 \le n \le 16$. Inset: behavior of the critical parameter 
$Z_{n}$ with respect to $n$. 
\vskip 0.25 in
\noindent


\begin{thebibliography}{99}

\bibitem{Rev} For a nice review on this fascinating topic, see 
McConnell, H. M. {\sl Annu. Rev. Phys. Chem.} {\bf 1991}, {\sl 42}, 171.

\bibitem{McC}McConnell, H. M. {\sl J. Phys. Chem.} {\bf 1990}, {\sl 94}, 4728.

\bibitem{Deu}Deutch, J. M.; Low, F. E. {\sl J. Phys. Chem.} {\bf 1992}, {\sl 96}, 7097.

\bibitem{Gol}Goldstein, R. E.; Jackson, D. P. {\sl J. Phys. Chem.} {\bf 1994}, {\sl 98}, 9626.

\bibitem{Lan}Langer, S. A.; Goldstein, R. E.; Jackson, D. P. 
{\sl Phys. Rev. A.} {\bf 1992}, {\sl 46}, 4894.

\bibitem{Thi}Thiele, A. A. {\sl Bell Syst. Technol.} {\bf 1969}, {\sl 48}, 3287.

\bibitem{McC4}McConnell, H. M.; de Koker, R.  {\sl J. Phys. Chem.} {\bf 1992}, {\sl 96},  7101.

\bibitem{Comment3}In theoretical models for ferrofluids and 
magnetic bubbles it is not necessary to introduce an {\it ad hoc} cutoff. 
In contrast to the strictly two-dimensional approach for monolayers, 
these systems are described as three-dimensional, presenting a finite 
thickness which prevents the integrals from diverging.

\bibitem{Mir}Miranda, J. A.; Widom, M. {\sl Phys. Rev. E} {\bf 1997}, {\sl 55}, 3758.

\bibitem{comment}See, for instance, the expressions calculated for 
$R_{n}$ in references~\cite{McC} (equation (30)) and~\cite{Deu} 
(equation (38)).

\bibitem{Mag}Magnus, W.; Oberhettinger, F.; Soni, R. P. {\it Formulas and Theorems for the Special Functions of Mathematical Physics}; Springer-Verlag: New York, 1966; Gradshteyn, I. S.; Ryzhik, I. M. {\it Table of Integrals, Series, and Products}; Academic Press: New York, 1994.

\bibitem{Rob}Robin L. {\it Fonctions Sph\'eriques de Legendre et Fonctions Sph\'eroidales}; Tome II and III; Gauthier-Villars: Paris, 1959.

\bibitem{Lee}Lee, K. Y. C.; McConnell, H. M. {\sl J. Phys. 
Chem.} {\bf 1993}, {\sl 97}, 9532.

\bibitem{McC2}McConnell, H. M. {\sl J. Phys. Chem.} {\bf 1992}, {\sl 96}, 3167.

\bibitem{Comment2}The reason for introducing such additional term into 
the line tension has its origin in the discussion of the 
applicability of Green's theorem to the calculation of the dipolar 
electrostatic energies of lipid domains. This issue is discussed 
in detail in reference~\cite{McC4}.

\bibitem{McC3}Keller, D. J.; Korb, J. P.; McConnell, H. M. {\sl J. Phys. 
Chem.} {\bf 1987}, {\sl 91}, 6417.

\bibitem{Mir2}Miranda, J. A.; Widom, M. {\sl Physica D} {\bf 1998}, {\sl 120}, 315.



\end{thebibliography}
\end{document}